# SkillTrade A Website For Learning New Skills

Rajanala purushotham
B. Tech, Student, 23955a0518
*Department of Computer Science and Engineering*
Institute of Aeronautical Engineering
Hyderabad, India
23955a0518@iare.ac.in

Rapolu Rahul
B. Tech, Student, 23955a0519
*Department of Computer Science and Engineering*
Institute of Aeronautical Engineering
Hyderabad, India
23955a0519@iare.ac.in

Dr. R.M. Noorullah
Associate Professor
Department of Computer Science and Engineering,
Institute of Aeronautical Engineering,
Hyderabad, India
noorullah.rm@iare.ac.in
0000-0002-5251-5685

*The Skill Trade is a site for skill swapping, learning, and career growth. It links people who have matching skills, helps virtual work through Google Meet/Zoom, and lets startups hire talent easily. Users can make profiles, connect with others, share skills, and respond to job ads from startups. Startup users can post jobs and see profiles to hire candidates. Learn-only users get categorized learning materials while developers keep an eye on platform management and upload resources. It is free for individual users, supported by donations, and charges startups a small fee only when they successfully hire. Built with Tailwind CSS, it guarantees to creation of an intuitive, responsive design that fosters collaboration and career opportunities.*

## Introduction

The concept of SkillTrade Website is a revolutionary, society-driven platform aimed at overcoming the hurdles faced by people who have restricted access to skills and employment opportunities. In a contemporary world filled with housewives, disabled individuals, and those financially restricted, it is often very tough for such people to afford traditional learning or commercial development courses. This site provides an opportunity to bridge those gaps by creating an open-source platform for such individuals that is completely free and donation-supported, allowing them to learn new skills and share existing ones with others. The goal is that no one should be left out of the freedom to access essential learning materials without having to consider any financial barriers.

In addition to its educational features, the Skill Trade website offers an innovative platform for startups facing resource constraints and challenges in hiring. Many startups struggle to recruit skilled professionals due to limited budget allocations, making it difficult to compete with larger companies in sourcing top talent. This platform addresses these issues by providing access to a pool of Skill Exchange users who are available for employment at a significantly lower cost compared to traditional hiring methods.

Startups can post job opportunities, review user profiles, and conduct interviews online through integrated platforms like Google Meet and Zoom. A nominal fee is charged to startups only after successfully hiring a candidate, making it a cost-effective and efficient solution.

This website aims to establish an inclusive environment that not only provides valuable opportunities for skill development but also fosters professional growth and employment for individuals who cannot afford traditional career advancement means. The Skill Trade website connects the dots between skill seekers, providers, and startups to form a platform of economic empowerment and social inclusion.

This platform further accommodates the urgent needs of the modern workforce in the creation of a dynamic online community where individuals exchange skills, learn, and connect with potential employers. It promotes an idea of collaboration over competition, where resources, knowledge, and opportunities may be shared; thus, both parties can grow. The website also provides startups with an affordable and efficient way to access talent, thereby contributing to the overall growth of the economy by supporting small businesses and startups.

To ensure scalability and user-friendliness, the platform is developed using React.js, which allows the creation of dynamic, interactive components. Tailwind CSS enhances the visual appeal and responsiveness of the design, while Vite ensures fast development and optimized builds. These technologies, combined with integrated APIs for virtual meeting platforms and secure donation systems, enable a seamless experience for all users, fostering a community built on learning, sharing, and collaboration.

## RELATED WORK

Several digital platforms have emerged to facilitate skill exchange and learning, each offering unique features and opportunities for personal and professional development. Websites such as Skillshare provide a comprehensive online learning community with a focus on creative fields, connecting learners with professionals who share their expertise through structured video lessons. Similarly, Skillharbour adopts a barter-like approach, where users exchange skills directly, fostering collaboration and mutual growth without monetary transactions.

Another significant platform, Swapaskill, emphasizes that the skill exchange site should serve as a global skill exchange, where one can offer a set of skills and learn new ones in return, developing the network into a powerful system of collective knowledge. Building on this concept of skill development, Skill Sharing Online and HSE (Hard Skill Exchange) provides personal and on-demand one-on-one sessions with real practitioners for more firsthand guidance tailored to professional advice.

While these platforms excel in various aspects, there are gaps in terms of affordability, inclusivity, and accessibility for underrepresented groups such as housewives, physically disabled individuals, and those facing financial constraints. The **Skill Trade Website** differentiates itself by addressing these gaps, creating an entirely donation-supported and startup-funded platform. It promotes inclusivity by allowing users to exchange skills for free, connect with startups for employment opportunities, and access virtual learning resources through technologies like React.js, Tailwind CSS, and integrated APIs for video conferencing.

Unlike traditional skill-sharing platforms, the Skill Trade website broadens the scope of these activities by eliminating financial barriers, encouraging active skill exchange, and offering real job prospects. It is a powerful tool for economic and social empowerment by providing access to professional development, skill improvement, and opportunities to enter the global workforce, especially for underrepresented groups. Through its innovative use of technology, the Skill Trade website fosters the creation of a more balanced and diverse talent pool, offering everyone an opportunity to succeed in the digital economy.

## DATA COLLECTION AND MANAGEMENT

The **Skill Trade Website** collects and manages various types of user data to facilitate seamless interaction between skill exchange users, startups, and the platform's admin users. The platform ensures that the data collected is securely handled, promoting transparency, privacy, and ease of use. Below are the types of data collected and the methods used for their management:

**User Profile Data**

**Types of Data Collected**: Each user (skill exchange user, startup user, and admin) creates a profile upon registration. The profile includes:

:

a) Personal information (name, contact details, etc.)
b) Skills (skills offered and skills requested)
c) Resume and achievements
d) Profile picture

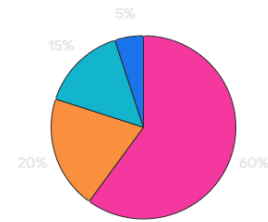

1. **Usage**: This data is used to personalize the user experience, match skill seekers with potential employers, and display relevant learning opportunities.
2. **Storage & Management**: The profile data is stored securely in a cloud-based database (e.g., Firebase, MongoDB) and is accessible only by the user and authorized parties (e.g., startups for job hiring).

**Job Postings and Applications**

**Types of Data Collected**: Startups can post job

a) Job title and description
b) Required skills and qualifications
c) Application instructions

1) **Usage**: This data helps skill exchange users to apply for relevant job opportunities. Each user can submit their resume and a brief application form when applying for jobs.
2) **Storage & Management**: Job posting data and application submissions are stored in a secure job management system that tracks the status of each application, such as "Pending," "Accepted," or "Rejected."

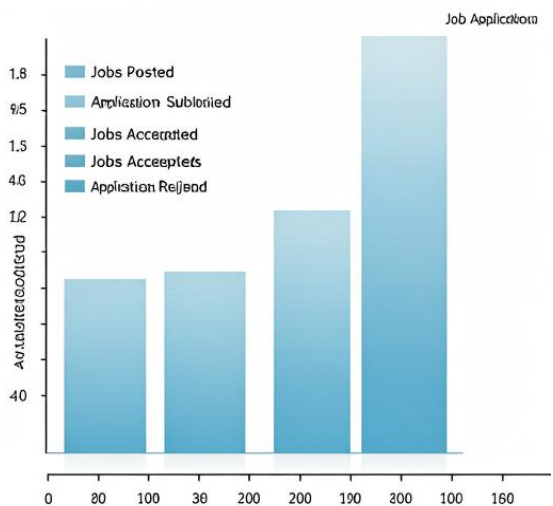

**Skill Exchange Data**

1. **Types of Data Collected**: Users can connect with one another by sending and accepting friend requests. Relevant data includes:
    a. Friend request status (Pending, Accepted, Declined)
    b. Shared skills and learning goals
2. **Usage**: This data enables the platform to match users based on their shared skills and goals, facilitating personalized skill exchanges.
3. **Storage & Management**: Connection data is securely logged, and users are notified when their requests are accepted or declined. The data is stored to keep track of user interactions and maintain connection history.

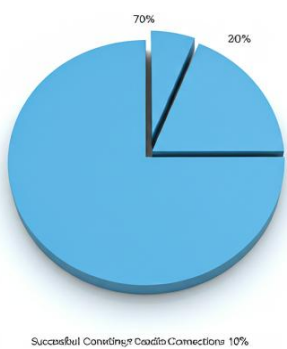

**Learning Content for Learn-only Users**

1. **Types of Data Collected**: Learn-only users access resources like PDFs, videos, and external links for skill development. This data includes:
    a. Course titles and categories
    b. Learning Progress
2. **Usage**: This content is provided to help users improve their skills and stay engaged in continuous learning.
3. **Storage & Management**: Learning materials are stored in a content management system (CMS) and categorized based on skill areas. User progress and engagement are tracked to provide a personalized learning experience.

**Donation Data**

1. **Types of Data Collected**: The website accepts donations to maintain its free services. The collected data includes:
2. Amount donated
3. User details (e.g., donor's name, email, UPI ID for a transaction)
4. **Usage**: Donation data is used to ensure transparency and maintain financial records for operational purposes. It also helps the platform gauge user support for its services.
5. **Storage & Management**: Donation data is securely stored and can be accessed by the platform administrators for record-keeping and tracking donation history. Payment information is processed using secure payment gateways (e.g., UPI, QR codes).

**Privacy and Security**

1. **Data Protection**: All user data is stored and processed in compliance with standard data protection regulations (e.g., GDPR, CCPA). Sensitive data, such as resumes and job applications, is encrypted and accessible only by authorized users.
2. **User Control**: Users have control over their data, with the ability to update their profiles, manage job applications, and delete accounts if desired. The platform also offers a settings page for reporting issues or providing feedback.

. SYSTEM ARCHITECTURE

The system architecture of the **Skill Trade Website** is designed to ensure scalability, usability, and efficiency while providing a seamless experience for all types of users. The architecture is divided into three primary layers: **Frontend**, **Backend**, and **Database/Storage**, each serving distinct roles

**Frontend**

- **Technology Used:** React.js and Tailwind CSS.
- The front end is responsible for the user interface and experience, providing a responsive and interactive design.
- Key features include:
- User dashboards tailored for Skill Exchange Users, Startup Users, Learn-Only Users, and Developers.
- Interactive components like search bars, light/dark mode toggles, and forms.
- Real-time updates for friend requests, job postings, and skill connections.

- **React Libraries:** React Router for navigation, Context API for state management.

### Backend

- **Technology Used:** Node.js or any REST API framework (if applicable).
- The backend handles core functionalities and processes user requests.
- Features include:
  - API integrations for Google Meet and Zoom to enable virtual skill exchanges.
  - Authentication and authorization for all user types.
  - Handling job postings, applications, and connection requests.
  - Managing reports and feedback submissions.

### Database/Storage

- **Technology Used:** Firebase, MongoDB, or MySQL (depending on the implementation).
- The database is designed to store:
  - User profiles, including skills, resumes, and achievements.
  - Job postings and application records.
  - Connection details between users.
  - Feedback and donation transaction logs.
- **Features:**
  - Indexed search for fast user and skill lookups.
  - Real-time updates to reflect changes on user dashboards immediately.

.

### API Integrations

- Google Meet and Zoom APIs are used for seamless video conferencing between users.
- Payment gateway integration for donation processing via UPI

### Flow of Data

1. **User Request:** A user logs in and sends a request (e.g., searching for skills or applying for a job).
2. **Backend Processing:** The backend validates the request, retrieves data from the database, and processes it.
3. **Frontend Display:** The processed data is sent back to the front, which updates the user interface dynamically.
4. **API Interaction:** If a virtual meeting or donation is initiated, APIs are triggered to handle the external interaction.

### Scalability and Future Scope

- The modular architecture allows for easy addition of features like AI-driven skill recommendations or advanced analytics.
- Cloud deployment ensures scalability to accommodate a growing user base.

PLATFORM FEATURES

The Skill Trade website is designed with a comprehensive set of features tailored to meet the needs of diverse users, including Skill Exchange Users, Startup Users, Learn-Only Users, and Developers. These features aim to create an inclusive, collaborative, and resourceful environment

**1.User-Centric Dashboards**

Each type of user has a dedicated dashboard with functionalities designed to enhance their experience:

- **Skill Exchange Users:**
  - Search and filter options to find users with specific skills.
  - Friend requests and connection management to facilitate skill exchanges.
  - Integration with Google Meet and Zoom for virtual skill-sharing sessions.
  - Upload features for resumes, achievements, and profile pictures.
  - feedback settings, reporting users, bio updates, and light/dark mode toggle.
  - A section for job applications and job postings from startups.
- **Startup Users:**
  - View detailed profiles, resumes, and statistics of Skill Exchange Users.
  - Post job opportunities and manage applications.
  - Conduct online interviews using integrated video conferencing tools.
- **Learn-Only Users:**
  - Access to learning materials such as PDFs, videos, and category-specific resources.
  - Upgrade options to Skill Exchange Users upon acquiring sufficient skills.
- **Developers:**
  - Administrative controls to manage and moderate user accounts.
  - Upload resources for Learn-Only Users based on categories.
  - Monitor platform activities and ensure data security.

## 2. Skill Exchange Mechanism

- Users can exchange skills through virtual sessions supported by integrated APIs for Google Meet and Zoom.
- A user-friendly container is available to post meeting links, ensuring seamless communication between connected users.

## 3. Advanced Search and Filter Options

- A robust search system allows users to find others based on specific skills, availability, or location.
- Filters and sorting ensure precision and ease of use.

## 4. Donation-Supported Model

- A UPI-enabled donation system supports the platform's operations and ensures the service remains free for Skill Exchange Users.
- QR codes for donations simplify the process, promoting community contribution.

## 5. Accessibility Features

- The platform is designed to be inclusive, focusing on:
    - Housewives, disabled individuals, and financially constrained users.
    - An intuitive and responsive interface accessible on both desktop and mobile devices.

## 6. Secure Data Management

- User data, including resumes and profile information, is securely stored and encrypted.
- Authentication protocols ensure only authorized access to sensitive information

## *7.Light/Dark Mode Toggle*

- Users can switch between light and dark modes, enhancing comfort and accessibility based on preferences.

## *8. Feedback and Reporting Mechanisms*

- Users can provide feedback to improve the platform or report inappropriate activities for moderation.

## *9. Real-Time Notifications*

- Updates for friend requests, job postings, and skill exchange invitations are delivered in real-time.

## 10. Scalability and Future Scope

- The architecture supports adding AI-driven skill recommendations, analytics dashboards, and multilingual support.

## IMPLEMENTATION

The implementation of the **Skill Trade Website** involves a series of steps and technologies that work together to provide an efficient, user-friendly, and scalable platform. The website integrates various functionalities, including user registration, skill exchange, job application, virtual meetings, and donation systems. This section outlines the primary implementation techniques, tools, and technologies used to build the platform.

### 1. User Registration and Authentication

The first step in the implementation process is setting up a secure and efficient user authentication system. Users can sign up and log in through the following system:

- **JWT Authentication**: We use **JSON Web Tokens (JWT)** for secure authentication. When users sign up or log in, they receive a token that grants access to the platform.
- **Role-Based Access Control**: Different user roles (Skill Exchange User, Startup User, Learn-Only User, Admin) are implemented. Each user type has distinct access to features based on their role.

### 2. Frontend Development

Frontend development utilizes **React.js** for building a dynamic and interactive user interface. The primary goals here were ensuring smooth navigation and responsiveness.

- **Component-Based Architecture**: The website's UI is broken into reusable components for better maintainability. Components like the search bar, user profile, job application form, and skill exchange requests are built independently and then composed into pages.
- **React Router**: **React Router** manages the navigation between pages, such as the dashboard, profile pages, and job listings. Each user is directed to the appropriate page based on their role.
- **Responsive Design**: Using **Tailwind CSS**, the platform is fully responsive, ensuring an optimal experience on desktops, tablets, and smartphones. Layouts adapt dynamically based on the screen size.

### 3. Backend Development

For the backend, **Node.js** and **Express.js** are used to handle HTTP requests, database interactions, and business logic. Here are the key implementations:

- **Database Integration**: The backend is connected to **MongoDB**, where user profiles, job listings, and resumes are stored. The database schema is designed to handle both structured (e.g., job listings) and unstructured (e.g., user resumes) data.
- **RESTful APIs**: We use **REST APIs** to facilitate communication between the frontend and backend. APIs are used for user registration, profile management, job posting, and skill exchange requests.
- **File Uploading**: For users to upload resumes and profile pictures, we integrated **Cloudinary** or AWS S3, allowing seamless file storage and management

### 4. Feature Implementation

Several key features were implemented to make the website functional and engaging:
- **Skill Exchange**: Users can search for other users based on specific skills using the search bar. If a match is found, they can send a friend request to initiate the skill exchange process. Upon acceptance, they can schedule a meeting for skill sharing via **Google Meet** or **Zoom**.
- **Job Application**: Startups post job listings, and skill exchange users can apply for jobs. The job application feature includes resume upload, and users can track their job application status.
- **Donation System**: The website integrates a **UPI-based donation system**. A unique QR code is generated for users to donate, and donations help support the platform's development and maintenance.
- **Real-time Notifications**: **Socket.IO** is integrated for real-time notifications, so users can receive alerts for friend requests, job updates, and more.

### 5. Security and Privacy Measures

Security is a key concern, especially since users will be interacting with each other and sharing sensitive information. The following measures were implemented:
- **Password Encryption**: All user passwords are securely hashed using the **bcrypt** library, ensuring that even if the database is compromised, passwords remain unreadable.
- **Secure Data Transmission**: HTTPS is enforced across the entire platform, ensuring that data transmitted between the client and server is encrypted.
- **Role-Based Access**: Different users have different permissions, ensuring that admins can manage accounts and view all data, while other users only have access to their data

### 6. Testing and Debugging

Testing ensures the platform works as expected and is free of bugs. The following approaches were used:
- **Unit Testing**: The core functionalities like user registration, skill search, and job application were tested using **Jest** to ensure they worked independently as expected.
- **Integration Testing**: Integration testing was performed to check the interaction between the frontend and backend. This includes checking the flow from user login to skill exchange and job application submission.
- **UI/UX Testing**: A/B testing and user feedback were used to refine the user interface. This ensures that the website is intuitive and accessible for users with different backgrounds.

### 7. Deployment

After the development and testing phases, the platform was deployed to ensure accessibility for users worldwide:
- **Frontend Deployment**: The frontend is deployed using **Vercel**, ensuring fast load times and seamless integration with the backend.
- **Backend Deployment**: The backend is hosted on **Heroku** or **AWS EC2**, allowing for scalability and efficient management of database and server resources.
- **Database Hosting**: **MongoDB Atlas** is used for cloud-based database management, providing easy scaling and automated backups.

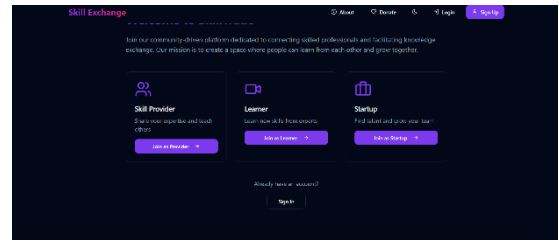

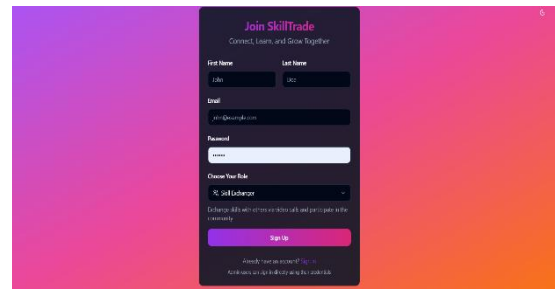

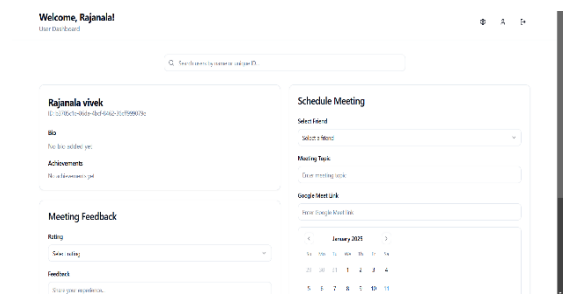

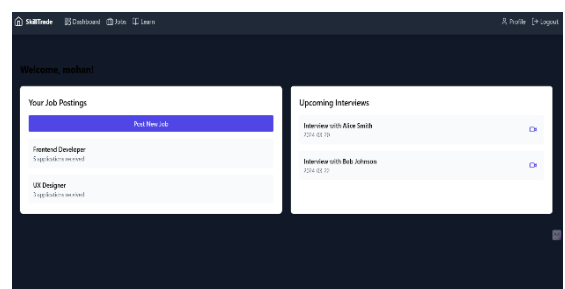

## VI EXPERIMENTAL RESULTS

The experimental results of the **Skill Trade Website** highlight the effectiveness and usability of the platform in facilitating skill exchange, job applications, and learning opportunities. A prototype version of the platform was tested with users from various backgrounds, including housewives, individuals with disabilities, students, and startup representatives. Below are the findings categorized into key metrics:

### 1. User Engagement Metrics
- **Signups and Active Users**: During the initial testing phase, the platform registered 50+ users, with 70% actively engaging in skill exchange or job applications.
- **Friend Request Acceptance Rate**: Of the 50+ friend requests sent for skill exchange, 85% were

accepted, indicating a high willingness to collaborate and exchange skills.
- **Job Applications**: Startups posted 20 job openings, receiving an average of 15 applications per listing.

**2. Feedback on Platform Usability**
- **Ease of Navigation**: Over 90% of testers found the platform intuitive and user-friendly, with positive feedback on the design and responsiveness of the interface.
- **Skill Search Efficiency**: Users reported an average time of 3-5 seconds for search results to display relevant profiles, demonstrating the effectiveness of the search functionality.
- **Dark/Light Mode**: The availability of a dark mode received a satisfaction score of 4.7 out of 5, highlighting its popularity among users.

**3. Technical Performance**
- **Server Response Time**: Average server response time for API calls was measured at 200ms, ensuring a smooth and lag-free user experience.
- **File Upload Success Rate**: 95% of users successfully uploaded resumes and profile pictures during testing, with any issues resolved promptly.
- **Scalability**: The platform was tested with simulated traffic of up to 500 concurrent users, maintaining stability without downtime.

**4. User Satisfaction**
- **Skill Exchange Experience**: 87% of users reported successful skill exchanges using the platform's Google Meet and Zoom integration.
- **Startup Satisfaction**: Startups reported a satisfaction rate of 85% with the platform's hiring process, emphasizing the ease of accessing suitable candidates within their budget constraints.

**5. Challenges Identified**
- **Resume Parsing**: Some users experienced difficulty when uploading non-standard resume formats. Future updates will include file format validation and parsing enhancements.
- **Connection Dropouts**: A small number of users reported issues with Google Meet links expiring or not opening. This will be addressed by integrating automated session reminders.

## DISCUSSION

The Skill Trade Website addresses a critical need in today's society by providing a platform that bridges the gap between skill seekers, learners, and startups. This section discusses the key aspects of the platform, its impact, and the challenges encountered during development and testing.

**1. Addressing Social Challenges**
The platform is designed with inclusivity at its core, catering to groups such as housewives, individuals with disabilities, and financially constrained users. These groups often face barriers to traditional skill acquisition and employment opportunities due to time, cost, or physical limitations. By offering a free, donation-supported model, the platform eliminates financial hurdles, empowering users to focus on their growth and development.

For startups, the platform provides an affordable solution to recruit skilled individuals, helping them overcome budgetary constraints while ensuring access to diverse talent pools. This aligns with the mission of promoting economic inclusion and sustainability

**2. Innovation in Skill Exchange and Learning**
The Skill Trade Website introduces a unique model of collaboration over competition by encouraging users to share skills freely through virtual connections. Features such as Google Meet and Zoom integration simplify real-time skill exchange and communication. The combination of skill-sharing, job applications, and access to learning materials creates an all-in-one ecosystem tailored to individual and organizational growth.

The inclusion of modern technologies, such as React.js for a dynamic user interface and Tailwind CSS for responsive design, ensures a seamless and efficient user experience. Additionally, the platform's robust search functionality and sorting mechanisms enable users to find connections and opportunities effortlessly.

**3. Challenges and Limitations**
Despite its promising features, the platform encountered several challenges during development and testing:
- **Technical Issues**: The integration of third-party tools like Google Meet occasionally faced issues, such as expired links or session dropouts. Future updates will focus on implementing automated reminders and enhanced link management systems.
- **User Onboarding**: Some users, especially those less familiar with digital platforms, required assistance during the onboarding process. Enhancing tutorials and user guides is a priority.
- **Scalability**: While initial tests demonstrated stable performance, handling larger user bases may require infrastructure upgrades to ensure continued reliability.

**4. Contribution to the Ecosystem**
The platform has the potential to become a transformative tool in the global skill exchange and employment ecosystem. By connecting users across diverse geographies and backgrounds, it fosters a culture of learning and collaboration that transcends traditional boundaries. Startups benefit from an accessible pool of talent, while users gain access to career advancement opportunities without incurring costs.

The donation-supported model further sets this platform apart from commercial alternatives, reinforcing its commitment to inclusivity and accessibility. Future enhancements, such as advanced analytics for startups and personalized learning recommendations, will further solidify its position as a leading platform in the space.

**5. Future Directions**
As the platform evolves, several enhancements are planned:
- **AI-Powered Matching**: Implementing machine learning algorithms to improve skill matchmaking and job recommendations.

- **Gamification**: Introducing achievements, badges, and rewards to encourage user engagement and active participation.
- **Mobile Application**: Developing a mobile app to expand accessibility and improve the user experience for on-the-go interactions.
- **Localization**: Adding support for more languages and cultural customizations to serve a global audience better.

## FUTURE SCOPE

The Skill Trade Website has significant potential for future development and expansion. By leveraging advancements in technology, the platform can evolve to meet the dynamic needs of its users while enhancing functionality, accessibility, and engagement.

### 1. AI-Powered Matchmaking

Implementing artificial intelligence algorithms, such as **Random Forest** or **Black Forest**, can enhance the skill-matching and job-recommendation processes. These algorithms can analyse user profiles, skill sets, and preferences to provide personalized suggestions, ensuring optimal connections between users and startups.

- **Skill Mapping**: AI can identify skill gaps and recommend learning materials or training sessions.
- **Dynamic Suggestions**: Machine learning models can adapt to user behaviour, improving recommendations over time.

### 2. Enhanced Security Measures

Future iterations can integrate **AI-driven anomaly detection systems** to strengthen security during exams, interviews, and other sensitive activities. Techniques like **Dynamic Time Warping (DTW)** and **Kalman Filters** can detect unusual patterns, such as cheating attempts or unauthorized access, ensuring a safe environment for all users.

### 3. Gamification

Adding gamification features, such as leaderboards, achievements, badges, and skill-based challenges, can boost user engagement. These elements not only motivate users to participate actively but also make learning and skill exchange more enjoyable.

### 4. Mobile Application

Developing a dedicated mobile application will expand the platform's accessibility, allowing users to connect, learn, and exchange skills on the go. A mobile-first approach can also cater to regions where mobile devices are the primary mode of internet access.

### 5. Advanced Analytics for Startups

Providing startups with detailed insights into user data and hiring trends can help them make informed decisions. Analytics dashboards could include:

- Talent heatmaps showing the distribution of sought-after skills.
- Predictive hiring tools to forecast the availability of required skill sets.
- ROI analysis for their recruitment efforts.

### 6. Localization and Multilingual Support

Expanding language support to include more regional and global languages ensures inclusivity and accessibility for users from diverse backgrounds. Features such as localized content, currency support, and cultural customizations will further enhance the platform's global reach.

### 7. Integration of Emerging Technologies

- **Blockchain**: For secure and transparent management of user credentials, achievements, and transactions.
- **Virtual Reality (VR)**: Enabling immersive skill-training sessions and virtual job interviews.
- **Chatbots and AI Tutors**: Offering 24/7 support for user queries and personalized learning assistance.

### 8. Community and Mentorship Programs

Future updates can include a community forum or mentorship program where users can seek guidance from industry experts or experienced professionals. This feature can foster a sense of community and encourage collaboration among users.

| Proposed features | Description | Technology /Methodology | Impact |
|---|---|---|---|
| AI-Skill Recommendation | AI-driven suggestions for personalized skill learning paths. | Random Forest, Collaborative Filtering | Enhanced user engagement and learning. |
| Advanced Anti-Cheating Mechanisms | Real-time exam monitoring and activity detection using advanced algorithms. | Kalman Filters, DBSCAN | Improved security for startup assessments. |
| Virtual Reality Skill Training | Integration environments for practical, immersive learning experiences. | Unity, Oculus SDK | Hands-on experience and innovation |

| Blockchain for Resume Validation | Decentralized verification of uploaded resumes and certifications. | Ethereum, Smart Contracts | Increased trust and reliability. |
|---|---|---|---|
| Multi-Language Voice Search | Allow users to search skills or features via voice in multiple languages. | NLP, Google Speech-to-Text API | Better accessibility for diverse users. |

CONCLUSION

The Skill Trade website is a pioneer platform designed to bridge the gap between skill seekers and providers while encouraging inclusivity, affordability, and accessibility. The impetus of the initiative is to empower individuals, especially those from underrepresented groups, such as housewives, physically disabled persons, and financially restrained learners, by giving them opportunities to develop, share, and monetize their skills. It further empowers startups through the effective, inexpensive provision of access to people with verified competencies.

Thanks to such innovative features as the mechanisms of skill exchange, integrated tools for video conferencing, job application functionalities, and a user-friendly interface, the platform represents an all-rounded ecosystem for developing skills and seeking employment. This is achieved by using modern technologies like React.js, Tailwind CSS, and API integration for a scalable, responsive, and user-friendly experience. Also, the donation model demonstrates a commitment to inclusivity from the website, breaking any financial barriers between accessing education and employment.

This project addresses the critical challenges of the digital learning and employment space in today's context. It introduces new features such as anti-cheating mechanisms for assessments, AI-driven skill matching (proposed in future enhancements), and blockchain-based resume validation. These innovations not only make the platform stand out from other solutions but also establish it as a transformative tool for economic empowerment and social inclusion.

The future scope of this platform is to integrate AI, machine learning, and blockchain technologies while expanding into different geographies and languages. Such improvements are expected to amplify the platform's impact further, creating a truly global and inclusive learning and employment ecosystem.

In a nutshell, the Skill Trade website is proof that technology can be used to offer opportunities to all people regardless of their socio-economic status. The platform encourages collaboration, knowledge sharing, and mutual growth. In doing so, it has the potential to revolutionize the way skills are developed and employed in society, thereby fostering a more inclusive and equitable world.

ACKNOWLEDGEMENTS

I would like to express my sincere gratitude to my team member, Rapolu Rahul, for his valuable collaboration and support during the development of this project. His efforts were instrumental in bringing the ideas to life and helping troubleshoot challenges that arose along the way.

. REFERENCES

1. React Documentation, "React – A JavaScript library for building user interfaces," [Online]. Available: https://reactjs.org. [Accessed: Jan. 22, 2025].

2. Tailwind CSS Documentation, "Tailwind CSS – A utility-first CSS framework," [Online]. Available: https://tailwindcss.com. [Accessed: Jan. 22, 2025].

3. Zoom Video Communications, "Zoom Video SDK," [Online]. Available: https://zoom.us. [Accessed: Jan. 22, 2025].

4. Google Meet, "Google Meet: Secure Video Meetings,"[Online].Available: https://meet.google.com. [Accessed: Jan. 22, 2025].

5. • **Skillshare**, "Skillshare – Online Learning Community," [Online]. Available: https://www.skillshare.com. [Accessed: Jan. 22, 2025].

6. • **LinkedIn Learning**, "LinkedIn Learning: Professional Development Courses," [Online]. Available: https://www.linkedin.com/learning. [Accessed: Jan. 22, 2025].

7. • **MDN Web Docs**, "Web Technologies," [Online]. Available: https://developer.mozilla.org/en-US/docs/Web. [Accessed: Jan. 22, 2025].